\documentclass[useAMS,usenatbib]{mn2e}

\usepackage{caption}
\usepackage{natbib}
\usepackage{graphicx}
\usepackage{textcomp}
\usepackage{latexsym}
\usepackage{varioref}
\usepackage{xspace}
\usepackage{makeidx}
\usepackage{verbatim}
\usepackage{tabularx}
\usepackage{epstopdf}
\usepackage{amsmath}
\usepackage{array}
\usepackage{rotating}
\usepackage{hyperref}
\usepackage{float}

\def\Kepler{\textit{Kepler}}
\def\RXTE{\textit{RXTE}}

\title[Reversibility of time series in XRBs and CVs]
{Reversibility of time series: revealing the hidden messages in X-ray binaries and cataclysmic variables}
\author[S. Scaringi et al.]
{S. Scaringi$^{1,2}$\thanks{E-mail: simo@mpe.mpg.de}, T.J. Maccarone$^{3}$ and M. Middleton$^{4}$\\ 
$^{1}$Instituut voor Sterrenkunde, K.U. Leuven, Celestijnenlaan 200D, B-3001 Leuven, Belgium \\
$^{2}$Max Planck Institute f{\"u}r Extraterrestriche Physik, D-85748 Garching, Germany\\
$^{3}$Department of Physics, Texas Tech University, Box 41051, Lubbock, TX 79409-1051, USA\\
$^{4}$Institute of Astronomy, Madingley Rd, Cambridge CB3 0HA\\
}

\begin{document} 

\date{}

\pagerange{\pageref{firstpage}--\pageref{lastpage}} \pubyear{2014}

\maketitle

\label{firstpage}

\begin{abstract}
We explore the non-linear, high-frequency, aperiodic variability properties in the three cataclysmic variables MV Lyr, KIC 8751494 and V1504 Cyg observed with \Kepler, as well as the X-ray binary Cyg X-1 observed with \RXTE. This is done through the use of a high-order Fourier statistic called the bispectrum and its related biphase and bicoherence, as well as the time-skewness statistic. We show how all objects display qualitatively similar biphase trends. In particular all biphase amplitudes are found to be smaller than $\pi/2$, suggesting that the flux distributions for all sources are positively skewed on all observed timescales, consistent with the log-normal distributions expected from the fluctuating accretion disk model. We also find that for all objects the biphases are positive at frequencies where the corresponding power spectral densities display their high frequency break. This suggests that the noise-like flaring observed is rising more slowly than it is falling, and thus not time-reversible. This observation is also consistent with the fluctuating accretion disk model. Furthermore, we observe the same qualitative biphase trends in all four objects, where the biphases display a distinct decrease at frequencies below the high-frequency break in their respective power spectral densities. This behaviour can also be observed in the time-skewness of all four objects. As far as we are aware, there is no immediate explanation for the observed biphase decreases. The biphase decreases may thus suggest that the fluctuating accretion disk model begins to break down at frequencies below the high frequency break.
\end{abstract}


\section{Introduction}

Aperiodic broad-band variability (including flickering) is observed throughout all classes of accreting compact objects: in X-ray binaries (XRBs, where the accretor is either a neutron star or a stellar-mass black hole), in Active Galactic Nuclei (AGN, where the accretor is a super-massive black hole), as well as in cataclysmic variables (CVs, where the accretor is a white dwarf). In general, most of the radiation produced by the accretion process is emitted through the release of gravitational potential energy (possibly excluding jet emission in low hard states of XRBs and classical nova outburst in CVs). In the case of accreting black holes (BHs) and neutron stars (NSs) the bulk of the emission takes the form of X-rays, whilst for accreting white dwarfs (WDs) most of the emission takes the form of optical/ultraviolet radiation. Because the size scales of white dwarfs (which are $\approx3$ orders of magnitude larger than NSs or BH event horizons for a similar mass) enforce larger characteristic radii on their accretion disks than for the disks around stellar mass BHs and NSs, they have cooler inner flows.  For the same reason, they also have longer characteristic timescales of variability (thermal, viscous and dynamical).

Previous studies of the aperiodic variability properties of accreting compact objects have shown how these systems all display similar power spectral densities (PSDs), which are well described by a combination of Lorentzians. In the case of XRBs the PSDs have high frequency breaks at around $0.1-1$Hz (\citealt{nolan81,vanderklis_aper,belloni00}), whilst similar PSD breaks in AGN are found at $\approx10^{-3}$Hz (\citealt{mchardy07}). Furthermore, the PSDs of CVs also show similar shapes, with high frequency breaks also occurring at $\approx10^{-3}$Hz (Scaringi~et~al.~2012a,2012b\nocite{scaringi12a,scaringi12b}). This similar phenomenology between all accreting compact objects might suggest a single origin for the aperiodic variability observed in the PSDs. 

Up until now one of the main tools employed to study time series in astronomy in general, and variability in compact accreting sources in particular, has been the amplitude of the Fourier transform, with very little emphasis on the phases. For periodic signals the magnitude alone can already yield a lot of information, especially when multiple periodicities or quasi-periodic oscillations (QPOs) are present in the data (see eg. \citealt{wijnands99,warner03}), but also for studying broad-band aperiodic noise (\citealt{heil14}). The phases of the Fourier transforms will also contain a wealth of information, which can be entirely complementary to the information contained in the magnitudes. One example is studying and understanding the phase difference between two simultaneously observed lightcurves in two distinct X-ray energy bands (\citealt{nowak99,uttley11,cassatella12a}), or at two distinct optical wavelength ranges (\citealt{scaringi13a}). The observed phase-lags between two lightcurves can be easily understood in terms of time-lags, and in the case of AGN and XRBs it can immediately be used to place constraints on the light-travel time between two different emitting regions. Optical reverberation mapping is also used to study the time-lags between emission lines versus continuum, and supports the idea of light travel time between an emitting and a reprocessing region as the source for the lags (\citealt{blandford82,gaskell88}). More recently attempts have also been made to study the non-linear coupling between two different frequencies in XRBs by looking at the bispectrum, and its related phases (\citealt{maccarone02,maccarone11,maccarone13}).

One of the main reasons why the phase information has not been exploited as much as the magnitudes is because it requires a large number of high quality independent measures of the Fourier spectrum (or some equivalent alternative). For studying XRBs, this has been achieved through the use of long observations with X-ray dedicated space-based missions such as \textit{RXTE}, \textit{XMM-Newton} or \textit{Chandra}, together with the relatively short intrinsic timescales of the systems (\citealt{vanderklis_aper} for a review). For CVs however this task requires much longer observations at optical wavelengths, which have to be quasi-continuous, with very little to no gaps in the time series. This is mainly due to the intrinsic timescales of CVs being three orders of magnitude slower in temporal frequency than XRBs. Such data is now becoming available thanks to the dedicated \Kepler\ optical timing mission, which allows observations of targets for many consecutive months on same cadence with very few interruptions, thereby providing the requisite evenly sampled data.

Throughout all of the aperiodic variability studies of AGN, XRBS and CVs over the last few years, one general picture has emerged, with the potential of describing all the observations across all systems. This is the fluctuating accretion disk model (\citealt{lyub,kotov,AU06}), which associates the observed variability to mass transfer rate fluctuations within the accretion disk. Each annulus in the disk produces its own aperiodic signature governed by the local viscous timescale, which increases in frequency towards the inner-most regions of the disk. As matter propagates inwards in the disk, the fluctuations which started further out in the disk couple, multiplicativley, with those generated further in. By the time the fluctuations reach the central emitting region they essentially contain information relating to the mass transfer rate fluctuations across most of the disk. One key ingredient of the fluctuating accretion model is that the accretion disk needs to possess a large scale height in order to retain the variability without it being damped (\citealt{churazov}). In this model, the fluctuations will start in the outer-most edges of the disk where the emissivity is low, and thus the contribution to the observed flux will also be low. As material moves inwards it will pass through regions where the emissivity will rise up to a maximum close to the compact object, and the infall time-scale will become shorter. Because of this transition through the disk we can thus expect the emission from the initial fluctuations to rise quasi-exponentially as they couple to the fluctuations generated at smaller disk radii. At some point the initial fluctuations will either reach the surface of the compact object or the event horizon of the black hole, causing a sudden end to the increase in emissivity, which in turn will result in a sudden drop in the emission profile. This will lead to noise-like flaring which is rising more slowly than it is decaying at frequencies close to the outer-disk edge viscous frequency.

A key to understanding and verifying this model further is to go beyond the simple amplitude studies of the Fourier transform, and to study the non-linear variability present in the time series. The first successful attempts of this have been: i) establishing that the lightcurves of the XRB Cyg X-1 are not time-reversible (\citealt{timmer00,maccarone02}), ii) that there exists a linear relation between the rms amplitude and flux of AGN, XRBs and CVs (known as the rms-flux relation, \citealt{uttley01}; Scaringi~et~al.~2012a;\nocite{scaringi12a} \citealt{heil12}) and iii) that there exists coupling between the variability components on all observable timescales for XRBs (\citealt{maccarone02,uttley05,maccarone11}). Additionally, under the assumption that the disk temperature increases at smaller disk radii, the propagation model is also able to qualitatively reproduce the observed hard X-ray lags in XRBs and AGN (\citealt{AU06,IK13}). Furthermore, if the assumption that the emissivity of the disk increase at smaller disk radii holds, another prediction of the propagating model is that the observed lightcurves should be composed of flares which are rising more slowly than they are decaying. This effect can be studied by using higher-order Fourier techniques such as the bispectrum which will be addressed here.

In this Paper we try to understand the temporal non-linear coupling observed within the time series of three CVs observed with \Kepler, and compare these to one XRB observed with \RXTE. The selected CVs are the two nova-like (NL) objects MV Lyr and KIC 8751494 as well as the dwarf nova (DN) V1504 Cyg, whilst the selected XRB is Cyg X-1. We study the non-linear variability by using the bispectrum (\citealt{hasselman,maccarone13}), and the related biphase and bicoherence, as well as the time skewness statistic (\citealt{pried}). These higher-order Fourier statistics, as well as the time skewness, will allow us to determine qualitatively if the shape of the noise-like flaring observed in the lightcurves of accreting compact objects match those predicted by the propagating fluctuations model. We provide in  Section~\ref{sec:bisp} an overview of both the bispectrum and the time skewness, and briefly describe their meaning and how they are calculated. In Section~\ref{sec:obs} we introduce the \RXTE\ dataset used for Cyg X-1, as well as the \Kepler\ dataset used for the three CVs considered in this study, namely V1504 Cyg, MV Lyr and KIC 8751494. The description of how we segment the various time series for the calculation of the bispectra and time skewness are also provided in Section~\ref{sec:obs}. Section~\ref{sec:results} presents our results on all four objects, whilst Section~\ref{sec:disc} provides a discussion of the results, with some possible explanations. Our conclusions are then drawn in Section~\ref{sec:conc}.

\section{The bispectrum}\label{sec:bisp}
Polyspectra are higher order time series analysis techniques analogous to the classical Fourier spectrum which take into account more than one Fourier frequency (\citealt{tukey84}). The bispectrum is the first of a series of polyspectra which can be used to understand phase correlations in single time series at two independent Fourier frequencies. More specifically, the bispectrum is related to two quantities in the time-domain: the skewness of a lightcurve in time and its asymmetry in flux. The bispectrum is defined for two frequency pairs as:
\begin{equation}
B(k,l)=\frac{1}{K} \sum_{i=0}^{K-1} X_i(k)X_i(l)X^*_i(k+l),
\label{eq:bispec}
\end{equation} 
where there are $K$ segments of equal length to a lightcurve, and where $X_{i}(f)$ denotes the Fourier transform of the $i$th segment, with the asterisk denoting its complex conjugate (see \citealt{mendel,fackrell} and references therein). It is a complex quantity which measures the magnitude and phase of the correlation between two independent Fourier frequencies. Although its variance will be affected by noisy signals, its value will be unaffected by Gaussian noise. Thus, a large number of independent lightcurves are required in order to produce reliable measurement of the bispectrum, each with high signal-to noise, and more importantly with stationary power spectra over the entire observation.

Given that the bispectrum is a complex number, it consists of a real and an imaginary component. The magnitude of the bispectrum has already been studied in previous astronomical lightcurves in the X-ray domain (\citealt{maccarone02,maccarone11}). With a particular normalization, it is referred to as the bicoherence, and is quite similar to the cross-coherence which is used to determine whether time-lags between two energy bands (or optical filters) are constant, and/or to determine the degree of frequency-dependent correlation in variability between two bands. It takes a value between 0 and 1, with 1 indicating there is total coupling between two Fourier frequencies, and 0 indicating no non-linear coupling. The most common form of the bicoherence used in astronomy is (\citealt{kim_powers}):
\begin{equation}
b^2(k,l) =
\frac{\left|\sum{X_i(k)X_i(l)X^*_i(k+l)}\right|^2}{\sum{\left|X_i(k)X_i(l)\right|^2}\sum{\left|X_i(k+l)\right|^2}},
\label{eq:bicoh}
\end{equation}
where $b^{2}(k,l)$ is the squared bicoherence (although \citealt{hinich} point out that other normalisations exist and are more sensitive to some types of non-linear behaviour). We note that the bicoherence can also be corrected for a bias term related to the number of segments used to compute it, as well as for Poisson noise (\citealt{uttley05}), but we do not apply these corrections here. The associated error on the bicoherence can be found in \cite{elgar}:
\begin{equation}
\rm{Var}[b^2(k,l)] = 
\left(\frac{4b^2}{2K}\right)\left(1-b^2\right)^3.
\end{equation}

The phase from Eq. \ref{eq:bispec} is called the biphase, and is defined as:
\begin{equation}
\hat{\beta}(k,l) = \rm{arg}[B(k,l)]
\end{equation}
in the interval $-\pi$ to $\pi$. It contains information relating to the skewness of the fluxes at specific Fourier frequencies as well as whether the time series is reversible. The errors on the biphases can also be found in \cite{elgar} as
\begin{equation}
\rm{Var}[\hat{\beta}(k,l)] = \left(\frac{1}{2K}\right)\left(\frac{1}{b^2}-1\right)
\end{equation}

It is important to note that the real component of the bispectrum describes the extent to which the flux distribution is skewed, whilst the imaginary component describes the extent to which the time series is symmetric in time (and thus reversible). Most accreting compact objects are shown to display log-normal flux distributions (\citealt{uttley05}; Scaringi~et~al.~2012a\nocite{scaringi12a}), which are positively skewed by definition and are a natural consequence of the propagating fluctuations model. Thus, it is important to realise that any case where the real component of the bispectrum is negative (or equally cases where the biphase lies between $\pi/2$ to $3\pi/2$) will be particularly interesting, as it would suggest that the flux distribution on a particular timescale would not follow the standard log-normality. 

The time-reversibility can be understood with a few simple examples. In cases where a periodic source displays pulses which rise more sharply than they decay (for example fast-rise slow decay profiles or FREDs) we would obtain a biphase of $-\pi/2$ at the periodic frequency, whilst a biphase of $\pi/2$ would suggest the opposite (i.e. pulses which rise more slowly than they decay). Phases of $0$ and $\pi$ on the other hand suggest symmetric, and thus reversible, pulses (see \citealt{maccarone13} for more illustrative examples). The biphase can also be used to study noise-like, rather than periodic signals, as will be done here. The interpretation of the biphase angle will be the same, however rather than looking at periodic pulses, the biphase of a noise-like signal will describe the shape of the lightcurve across a broad range of frequencies. 

The asymmetry in lightcurves can also be analysed in the time-domain rather than the frequency-domain using the time skewness statistic, which is similar but not identical to the biphase as it considers only a single characteristic time-scale. The time-skewness was developed and applied for the first time by \cite{pried}. It has been slightly modified and applied to \RXTE\ lightcurves of XRBs by \cite{maccarone02}, who obtained the first evidence for asymmetry in XRB time series using this statistic. It is defined as:
\begin{equation}
\begin{split}
TS(\tau) = \frac{1}{\sigma^3}\frac{1}{K} \sum_{i=0}^{K-1} (s(t)-\bar{s})^2(s(t-\tau)-\bar{s}) \\
- (s(t)-\bar{s})(s(t-\tau)-\bar{s})^2
\label{eq:TS}
\end{split}
\end{equation}
where $\tau$ is defined to be $u$ times the time bin size, $s_i$ is the count rate in the $i$th element of the lightcurve, and $\sigma$ the standard deviation of the whole lightcurve. In a similar manner to the biphase, a negative time skewness on a specific timescale would suggest that the noise-like features in lightcurve are best described as rising more sharply than the fall, and viceversa for positive time skewness.

\section{Observations and data reduction}\label{sec:obs}
Here we will describe the data used in this work as well as the data analysis procedures adopted. We will first describe the \textit{RXTE} data analysis procedures for the XRB Cyg X-1 and then describe those of the three CVs with data obtained from \Kepler. This study is mostly concerned with the high-frequency variability found in these objects. This is mainly because i) the PSDs are most stationary at high frequencies and ii) studying the high frequency variability allows to segment the time series more finely, thus providing more segments for averaging and resulting in smaller errors on the bispectral analysis. 

\subsection{Cyg X-1}
The XRB studied in this work is Cyg X-1 during a hard state. The data for Cyg X-1 is the same as that used in \cite{maccarone00} and \cite{maccarone02}, and is taken from the \RXTE\ proposal 30158, which includes photons detected across the full \RXTE\ energy range. This consists of 12 observations made in December of 1997.  We use Standard 1 standard products light curves from HEASARC\footnote{\url{http://heasarc.gsfc.nasa.gov/}} for these data sets, giving us the full range of \RXTE\ energies with 0.125 second time resolution. A total of 37171 seconds of data are used. We have chosen to probe the high frequency variability above $\approx10^{-1}$Hz, and we thus segment the time series into 2951 equal length segments, each of 12.5 seconds long. We computed the PSD of each segment individually, and averaged all to obtain the time averaged rms normalised PSD (in units of (rms/mean)$^2$/Hz, see \citealt{miyamoto,belloni02}). We further computed the time skewness for each segment individually, and averaged all to obtain the time averaged $TS$, whilst the errors were obtained from the rms variations of the individual measurements used. Finally we also computed the bispectrum and related biphase and bicoherence as described in Section~\ref{sec:bisp}.

\subsection{CVs}
All lightcurves of CVs used in this work have been obtained by the \Kepler\ satellite. Since we are interested in comparing the biphases of the high frequency variability we only study sources which have short cadence (SC, $58.8$ seconds) data available. All lightcurves were obtained from the Mikulski Archive for Space Telescopes (MAST) in reduced and calibrated format after being run through the standard data reduction pipeline (\citealt{jenkins}). Only Single Aperture Photometry (SAP) is considered here.

We identified 10 CVs observed in SC mode by \Kepler. Only three of these, however, have high enough signal-to-noise ratios to not be dominated by Poisson noise at the highest observable frequencies. The systems selected are the two nova-like variables KIC 8751494 and MV Lyr as well as the dwarf nova V1504 Cygnus. Table \ref{tab:1} shows a log of the available \Kepler\ data for these sources. All three sources have already been subject to previous studies with \Kepler\ data (\citealt{kato13}; Scaringi~et~al.~2012a,2012b\nocite{scaringi12a,scaringi12b}; \citealt{scaringi14,cannizzo12}). 

\begin{table}
\centering
\begin{tabular}{l l l}
 \hline
 Name & Quarters & Total time (days) \\
 \hline
 \hline
 KIC 8751494  & 2,3,5,16    &  220.4   \\ 
 MV Lyr  &  14-17  &  273.2   \\   
 V1504 Cyg   &  2-14 & 809.5 \\  
  \hline
\end{tabular}
\caption{Log of \Kepler\ data for the 3 CVs studied in this work. For KIC 8751494 the whole available \Kepler\ SC data has been used. Although MV Lyr has been observed in SC mode throughout quarters 2-17, this work only considers data between quarters 14-17 ($56107-56424$ JD). Furthermore, the total time listed for V1504 Cyg is only the time used in this work after the removal of the normal- and super-outbursts. }
\label{tab:1}
\end{table}

It is important that the PSDs of all three objects are stationary during the observations (\citealt{vaughan03}) to obtain reliable measurements for both the bispectrum and the time skewness. Because the lightcurve of V1504 Cyg contains many normal and super-outbursts (\citealt{cannizzo12}) we cannot simply use the whole dataset, but instead use only the quiescent intervals for this study in order to maintain stationarity throughout the observation. We have adopted a conservative method to only select quiescent intervals in the \Kepler\ lightcurve of V1504 Cyg using a combination of gradient changes within the lightcurve, together with visual inspection to remove all outbursts. For MV Lyr a previous study has shown that the PSD is in fact quasi-stationary at frequencies above $\approx5\times10^{-4}$Hz (Scaringi~et~al.~2012b\nocite{scaringi12b}). We have employed a conservative cut, and only select data in between $56107-56424$ JD (see Fig.~1 of \citealt{scaringi14}), where the lightcurve of MV Lyr is also stationary at frequencies below $5\times10^{-4}$Hz. On the other hand KIC 8751494 does not display any sudden flux changes, and we thus use all available \Kepler\ data for this source. We checked for PSD stationarity in all three sources by looking at the PSDs obtained from small \Kepler\ data segments to ensure the shape of the PSDs are always consistent within errors. 

For all 3 CVs we decided to probe the high frequency variability above $10^{-4}$Hz, which ensures no contamination from the orbital or superhump signals in these systems. We thus segmented all lightcurves into 100 minute long segments. After avoiding any data gaps in the lightcurves\footnote{Data gaps occasionally occur due to \Kepler\ entering anomalous safe modes. Additionally data gaps in V1504 Cyg occur due to the removal of outbursts.}, this resulted in 3122, 3857 and 11210 segments for KIC 8751494, MV Lyr and V1504 Cyg respectively. We then computed the PSD for each segment individually and averaged all PSD to obtain the averaged rms-normalised PSD (\citealt{miyamoto,belloni02}). Additionally for each source we computed the bispectrum, bicoherence and biphase, together with the associated errors. We also computed the time skewness, $TS$, for each segment individually, and averaged all to obtain the mean $TS$. The errors on $TS$ were obtained from the rms variations of the individual measurements used in the averaging, and may be overestimated if there exists intrinsic variations in the time skewness.

\section{Results}\label{sec:results}
\begin{figure*}
\includegraphics[width=0.24\textwidth, height=0.6\textheight]{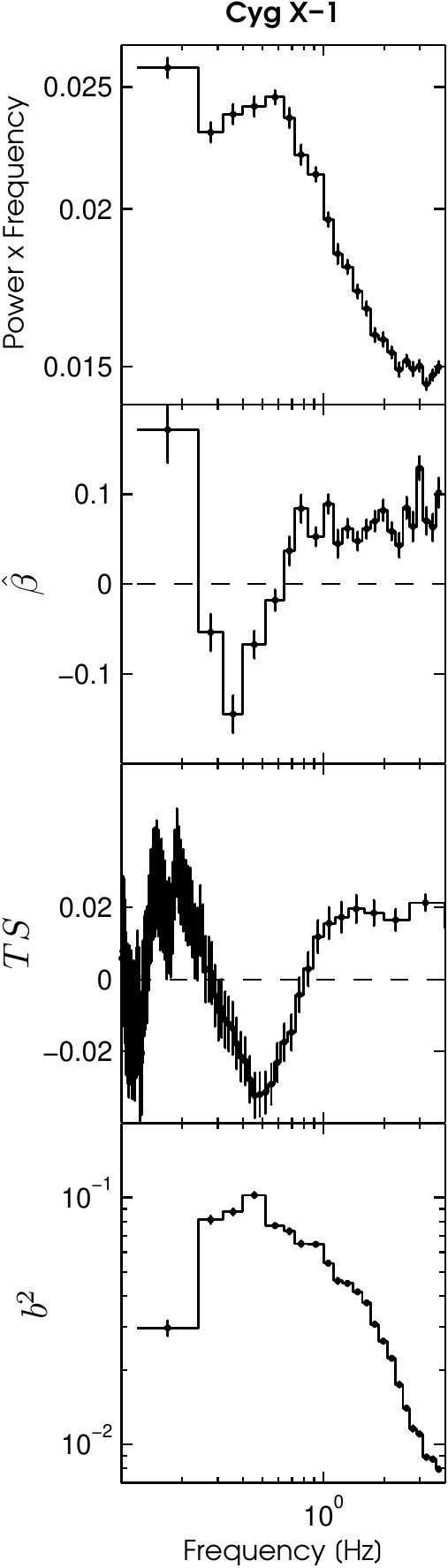}
\includegraphics[width=0.24\textwidth, height=0.6\textheight]{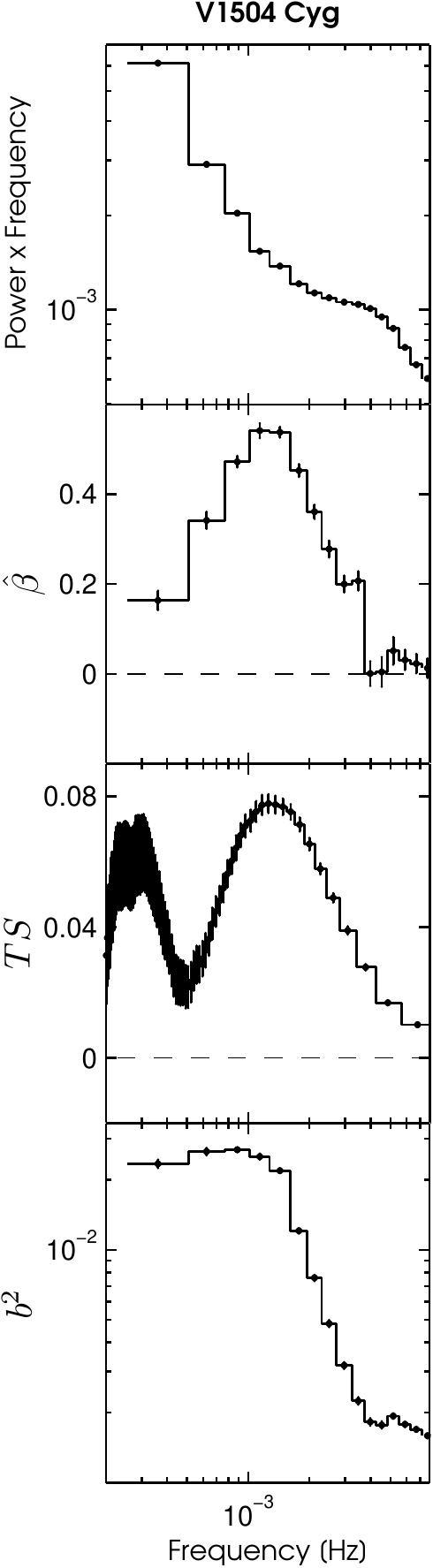}
\includegraphics[width=0.24\textwidth, height=0.6\textheight]{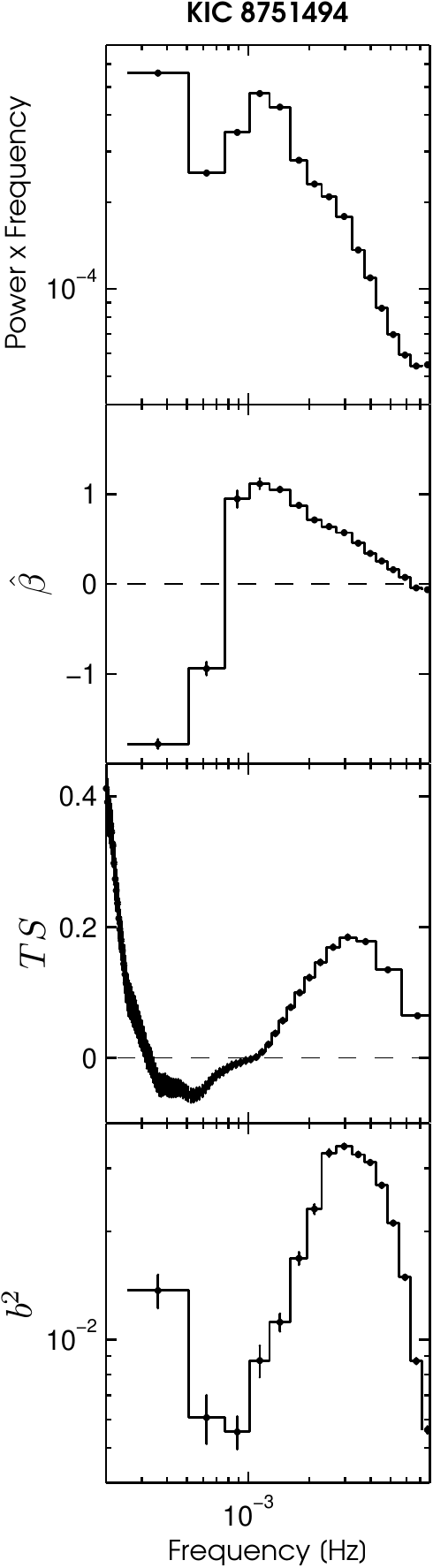}
\includegraphics[width=0.24\textwidth, height=0.6\textheight]{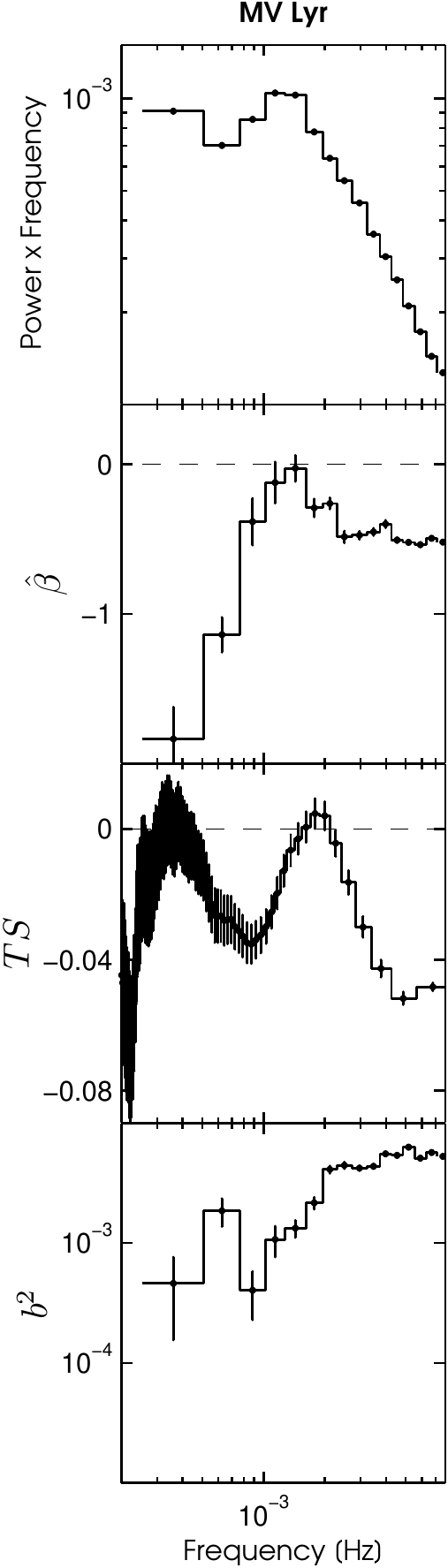}
\caption{Results from our Fourier, bispectral and time skewness analysis. From left to right each column displays the results for Cyg X-1, V1504 Cyg, KIC 8751494 and MV Lyr respectively. In each row from top to bottom are the respective rms-normalised PSD (Poisson noise not subtracted), biphase, time skewness and bicoherence. Although the time skewness is calculated in the time domain, here we plot it in units of $1/$time in order to allow easy comparison with the other plots. We note that the biphase and bicoherence have summed frequencies on the x-axis, as described in \ref{sec:results}.}
\label{fig:1}
\end{figure*}

Since the biphase and the bicoherence are a function of two independent Fourier frequencies, it is natural to display three-dimensional plots, or two-dimensional contour plots. However, we find from inspection that the three-dimensional landscape for both do not show any individual peaks, but rather smooth trends. For clarity and display purposes, we have averaged the biphase and bicoherence for each source along constant frequency diagonals ($f(k)+f(l)=$constant), such that the biphase/bicoherence matrices are collapsed into one dimension. The errors on both have then been computed through standard error propagation. Furthermore, all plots except for the time skewness have been geometrically binned in frequency with a $1.05$ binning factor. Our results for the PSDs, biphases, time skewness and bicoherence are shown in Fig.~\ref{fig:1}. Each column in Fig.~\ref{fig:1} shows the results for each source individually. Note that the frequency range probed is the same for all CVs, but is $\approx3$ orders of magnitude higher for the XRB Cyg X-1. Although the time-skewness has been calculated in the time domain, we plot our results in units of $1/$time in order to allow easy comparison with the other plots.

The first thing to note from Fig.~\ref{fig:1} are the shapes and frequency ranges of the PSDs. The obtained PSD for Cyg X-1 is consistent with the hard state PSD obtained in previous studies (\citealt{pott,gleiss04}). Also, the PSD of MV Lyr is consistent with previous \Kepler\ studies for this object (Scaringi~et~al.~2012a,2012b\nocite{scaringi12a,scaringi12b}; \citealt{scaringi14}). On the other hand we display the PSDs of V1504 Cyg and KIC 8751494 for the first time here. Qualitatively speaking, KIC 8751494 has a very similar PSD shape to MV Lyr, with a similar break frequency at $\approx10^{-3}$Hz. Closer inspection of KIC 8751494 however reveals a second break at $\approx3\times10^{-3}$Hz, which will be examined in more detail in a future paper with respect to the fluctuating accretion disk model as has already been done for MV Lyr (\citealt{scaringi14}). The DN V1504 Cyg on the other hand has a very different PSD shape when compared to the NLs MV Lyr and KIC 8751494, with a high frequency break at $\approx5\times10^{-3}$Hz and a lower break at frequencies below $\approx10^{-4}$Hz. NL CVs are thought to be in a state somewhat comparable to the hard state of XRBs, where they posses an optically thin (\citealt{pratt}), possibly extended (\citealt{scaringi14}), boundary layer. This distinction might well be the reason why the PSDs of the NL shown in Fig.~\ref{fig:1} resemble the PSD of Cyg X-1 in the hard state, and why the DN V1504 Cyg displays a qualitatively different PSD shape. Thus the distinction between the two different types of PSDs can potentially be attributed to the different nature of the sources (DN vs. NL), however a detailed comparison and interpretation is beyond the scope of this paper.

The biphases for all sources are shown in the second row of Fig.~\ref{fig:1}. All obtained biphases at all frequencies are below $\pi/2$ in amplitude, suggesting that the flux distributions are positively skewed on all observed frequencies. Furthermore, the biphases of all CVs display a decline at frequencies below $\approx10^{-3}$Hz, whilst Cyg X-1 shows a decline at frequencies below $1$Hz, followed by a biphase ``flip'' at the lowest frequencies. All four sources do, however display the same qualitative behaviour, where the biphases commence a decline at frequencies below the high frequency break seen in the PSDs. The same qualitative behaviour can also be observed in the time skewness plots shown in the third row of Fig.~\ref{fig:1}, where the time skewness is seen to drop and reach a minimum at the point where the corresponding PSDs show a decline in power below the high frequency break. The only source where this qualitative behaviour is potentially ambiguous is the DN V1504 Cyg, where the PSD does not allow us to determine where exactly the high frequency broad Lorentzian component begins to lose power at low frequencies. Nevertheless, all other systems display a biphase and time skewness decrease at the point where the high frequency Lorentzian begins to lose power; it is also possible that this is happening in V1504 Cyg, but that we cannot observe this in the PSD. If this were the case, the biphase would allow us to place tighter constraints on the width of the high frequency broad Lorentzian observed, even when this component is overpowered by lower frequency ones.

It is not easy to understand how the biphases and time skewness plots relate to the noise-like variability observed in the lightcurves. However, taking insight from the biphase study of periodic modulations of \cite{maccarone13}, we can deduce that the shape of the noise-like variability is changing as a function of frequency. Specifically from Fig.~\ref{fig:1}, we see that for all four sources, the noise-like flaring will be rising more slowly than it is decaying at frequencies where their respective PSDs break at high frequencies. This qualitative behaviour then starts to reverse at the lowest frequencies, such that the flare shapes become either more symmetric, or tend to display flares which rise faster than they decay. This qualitative behaviour will be further discussed in Section~\ref{sec:disc}.

Finally, the bicoherence for all sources is shown in the fourth row of Fig.~\ref{fig:1}. All sources have non-zero bicoherence, which indicates that there exists non-linear coupling between all timescales probed. This in turn supports the idea that the variability is produced within the accretion disk, and that the accretion-induced variability is coupled multiplicativley as described by the fluctuating accretion disk model. We note that the bicoherence results for Cyg X-1 display somewhat larger values than those reported in \cite{maccarone02} and \cite{uttley05}. This is because the data segments used here are shorter than those used previously, which affect the absolute bicoherence normalisation (\citealt{hinich}). Increasing the time segments to match those used previously would make our results consistent with previous studies.

\section{Discussion}\label{sec:disc}

The biphase and time skewness of both XRBs and CVs appear to display the same qualitative behaviour at frequencies around where their respective PSDs display high-frequency breaks. Firstly, all biphase amplitudes are found to be smaller than $\pi/2$ on all timescales studied here. This implies asymmetric flux distributions which are positively skewed, with long tails at high fluxes, consistent with the log-normal flux distributions observed in a number of accreting compact objects (\citealt{uttley01,uttley05}; Scaringi~et~al.~2012a\nocite{scaringi12a}), and consistent with the fluctuating accretion disk model (\citealt{AU06}). Secondly, all objects display positive biphases at frequencies where their corresponding PSDs break at high frequencies, suggesting that the noise-like flaring observed is rising more slowly than it is decaying. These observations are consistent with the fluctuating accretion disk model with an increasing emissivity towards smaller disk radii. This is caused by the outer-most generated fluctuations in the disk travelling through regions of higher emissivity and faster infall time-scales up to the surface or event horizon of the compact object, where the emissivity suddenly drops. We also note that the underlying variability process still needs to be intrinsically non-linear \textit{before} modification by the emissivity, for the model to produce non-zero bicoherence and hence a biphase. Thirdly, the biphases of all four objects studied here display a distinct biphase decrease at frequencies below the high-frequency break in their respective PSDs. As far as we are aware of there is no immediate explanation for the observed biphase decreases. However we note once again that this decrease occurs where the aperiodic variability begins to lose power at low frequencies (or flattens), and might in fact suggest similar physical changes are occurring at these specific frequencies in both XRBs and CVs. 

One way to explain this is in terms of the accretion disk geometry and state. Up until now, the standard fluctuating accretion disk model has been successful in reproducing the high frequency aperiodic variability in accreting compact objects, including qualitatively predicting the rms-flux relation, Fourier-dependent hard  lags, as well as the location and shape of the high-frequency broad Lorentzian component. However, it has not yet been successfully applied to the very low frequency variability, including the very low frequency broad QPO(s) observed in several XRBs ($<10^{-1}$Hz, \citealt{pott}), as well as CVs (Scaringi~et~al.~2012b\nocite{scaringi12b}). It could well be that the various broad low-frequency QPOs are being generated at different locations in the accretion disk, and that the disk itself might exhibit different geometries/states simultaneously at different radii. This scenario is not new, and has often been invoked to explain state changes in XRBs, where the interplay between a geometrically thin (optically thick) and a geometrically thick (optically thin) disk at different radii are responsible for the X-ray spectral changes during the outbursts (\citealt{done07}). 

More specifically, the properties of the fluctuating accretion disk model can only reproduce the geometrically thick inner disk, which is responsible for the high frequency broad Lorentzian component observed in CVs and XRBs (\citealt{ID10,ID11,ID12,scaringi14}). Although from the model the high frequency break strongly depends on the emissivity index (but to a lesser extent on the viscous timescale at the inner-most edge of the disk and the mass of the central compact object), the low frequency turnover of the broad high-frequency Lorentzian can be shown to be insensitive to all model parameters except for the viscous timescale at the outer, geometrically thick, disk edge. For XRBs the outer-edge of this component is inferred to be $\approx 25R_{g}$ during bright hard states (where $R_g$ is the gravitational radius, see \citealt{ID12,IK13,uttley11}), whilst from the modelling of one CV (MV Lyr, \citealt{scaringi14}) this radius is inferred to be $\approx0.1R_{\odot}$.

In the case of Cyg X-1, V1504 Cyg and KIC 8751494, the biphases are positive at frequencies close to the PSD break, which is consistent with the general propagating model picture. Only at frequencies below the break do the biphases start to decrease, and this is also where the current model breaks down. Thus it is possible that the biphases below the PSD break frequency are related to the termination of the the inner geometrically thick disk, and the start of the disk which is dominated by the geometrically thin component. One other possible explanation could be related to the boundary conditions at the outer-edge of the inner disk. Because, mass transfer is thought to be constant in the outer geometrically thin disk, it is possible that, at the transition to the geometrically thick disc, an increase in mass transfer (and thus flux) would be followed by a ``waiting time'' before the mass transfer can become large again due to a reservoir effect. MV Lyr is the only object where the biphases are close to $0$ close to the PSD peak, and this could suggest a potential problem with the propagating model applied to this source. We note, however that the same decrease in biphases, as seen in other sources, is observed in MV Lyr as in the other sources, and also occurs at frequencies where its corresponding PSD begins to lose power. Thus qualitatively, MV Lyr also seems to follow the same trend as the other sources.

It is clear from the above that detailed modelling of the low frequency variability is required in order to be able to explain all variability properties in a consistent manner. Such modelling is beyond the scope of this paper, but we expect that any attempt would have to include the contribution of the geometrically thin outer disk, which up until now has been somewhat neglected in the propagating fluctuations model. The addition of this component might help explain the observed biphase trends presented here, and might also help explain both the soft X-ray lags observed in XRBs and AGN (\citealt{uttley11,demarco}), as well as the soft optical lags observed in CVs (\citealt{scaringi13a}). In XRBs and AGN the soft lags are simply interpreted as the light-travel time form the hard X-ray emitting corona onto the soft X-ray emitting disk. This is also supported by X-ray observations of the broad iron L line emission lagging the X-ray continuum (\citealt{zoghbi10,kara13}). For CVs however the observed time-lags are too long to be explained solely by this mechanism, and an additional reprocessing timescale must be included for this model to work. The disk in CVs could potentially absorb the high energy photons originating from the inner geometrically thick disk, and reprocess them on the thermal timescale. As long as the thermal timescale is shorter than the observed aperiodic variability, the fluctuations produced in the geometrically thick disk will not be damped during reprocessing, and will modulate the optical bands as well. 

Finally we point out that this study has been based on only one XRB and only three CVS (two NLs and one DN). More systems are therefore required (including more of those studied here but also AGN and NS accretors which will be the subject of a future paper in preparation) to provide a more general picture of the phenomenological bispectral properties of accreting compact objects. If the biphases were found to consistently drop at frequencies corresponding to where the PSDs would lose power below the high frequency breaks, then this could be providing further evidence for a universal model to explain aperiodic variability across all compact accretors. In particular, the biphase statistic might have the potential to disentangle the geometrically thick-to-thin transition region of the accretion disk, an understanding of which is important for QPO timescales

\section{Conclusion}\label{sec:conc}
We have compared the non-linear, high-frequency, aperiodic variability of the three CVs MV Lyr, KIC 8751495 and V1504 Cyg observed with \Kepler, and the XRB Cyg X-1 observed with \RXTE. This comparison has been done using the higher order Fourier bispectrum and its related biphase and bicoherence, as well as the time-domain time-skewness statistic. We find that the PSD shapes of the two NL CVs MV Lyr and KIC 8751495 are qualitatively similar to the hard state PSD of Cyg X-1, all displaying a high frequency broad Lorentzian component breaking at $\approx10^{-3}$Hz in the CVs and at $\approx5\times10^{-1}$Hz in Cyg X-1. The PSD of the DN V1504 Cyg (exclusively obtained during its quiescent intervals) on the other hand displays two district breaks, one below $10^{-4}$Hz and one at $4\times10^{-3}$Hz. The difference between the PSDs of the two NL and the DN might be attributed to the different states of these objects, but a detailed comparison and modelling is beyond the scope of this paper. We further establish that all biphases at all frequencies observed are smaller than $\pi/2$, which suggests that the flux distributions are positively skewed on all frequencies, consistent with previous observations (\citealt{uttley01,uttley05}; Scaringi~et~al.~2012a\nocite{scaringi12a}), as well as with the fluctuating propagation disk model (\citealt{AU06}). Additionally, the biphases of all objects, except MV Lyr, display positive values at frequencies where their respective PSDs break at high frequencies, suggesting that the noise-like flaring observed is rising more slowly than it is decaying. This is also consistent with the fluctuating accretion disk model, where the outer-most generated fluctuations in the disk are travelling through regions of higher emissivity and faster infall time-scales as they move inwards, up to the surface or event horizon of the compact object where the emissivity suddenly drops. Finally, and more importantly, all objects are shown to display very similar qualitative trends in their biphases, with a distinct biphase decrease at the point where their respective PSDs begin to lose power at frequencies below the high frequency break, which can also be observed in their respective time-skewness plots. The only source where this qualitative behaviour is potentially ambiguous is the DN V1504 Cyg, as the PSD does not allow us to determine where exactly the high frequency broad Lorentzian begins to lose power at low frequencies. As far as we are aware, there is no immediate explanation for the observed biphase decreases. One potential explanation for this could be related to the accretion disk geometry and state, where the biphase decrease marks the outer-edge of the inner geometrically thick accretion disk, and possibly the transition to the outer geometrically thin disk. It could also be that the biphase decrease observed at low frequencies is related to the boundary conditions at the outer-edge of the inner disk, where the mass transfer at the inner disk outer-edge is followed by a ``waiting time'' as a consequence of a reservoir effect.

The current propagating fluctuations model is only able to reproduce the high frequency variability observed in accreting compact objects with a geometrically thick inner hot flow, and the biphase decreases observed here could potentially suggest that this model begins to break down at frequencies below the peak of the high frequency broad Lorentzian component. It is clear that any real attempt at interpreting the non-linear Fourier behaviour observed here will require detailed modelling of the accretion flow, but we do note that any modelling attempt would require the inclusion of the geometrically thin-to-thick transition in order to explain the observations self-consistently. Finally, the qualitative non-linear Fourier similarities between CVs and XRBs presented here, together with their qualitatively similar PSD shapes and Fourier-dependent time-lags, suggest that a similar physical mechanism generating the broad-band variability is at play across all accreting compact objects, irrespective of their mass, size or type. Thus, any plausible physical model seeking to explain this variability will have to explain the observations across all types of accreting compact objects.

\section*{Acknowledgements}
This research has made use of NASA's Astrophysics Data System Bibliographic Services. S.S. acknowledges funding from the FWO Pegasus Marie Curie Fellowship program, as well as thanking the Max-Planck-Institute for Extraterrestrial Physics for the long research visit. S.S. also wishes to thank Texas Tech University for the hospitality during the period when this project started. Additionally this work acknowledges the use of the astronomy \& astrophysics package for Matlab (Ofek in prep.). The authors would also like to thank the referee Phil Uttley for the usefull comments which have imporved the paper.

\bibliographystyle{mn2e}
\bibliography{bisp_paper}


\end{document}